\documentclass[useAMS,usenatbib,useverbatim]{mn2e}

\usepackage{graphicx}
\usepackage{psfig}

\bibpunct{(}{)}{;}{a}{}{,}    
\usepackage{amsmath}
\usepackage{amssymb,graphicx}
\usepackage{verbatim}

\usepackage{journal_names}
\usepackage{epstopdf}
\usepackage{lscape}

\title{A comprehensive long term study of the radio and X-ray Variability of NGC~4051 Paper II}

\author[S.Jones et al]{S.Jones$^{1}$\thanks{E-mail:
sadie.jones@soton.ac.uk (SJ)}, I.M$^{\mbox{c}}$Hardy$^1$, T.J. Maccarone$^2$\\
$^1$Department of Physics \& Astronomy, University Southampton, SO17~1BJ, UK \\
$^2$Department of Physics \& Astronomy, Texas Tech University, PO Box 40151, Lubbock, TX 79409-1051, USA\\}

\date{Accepted 2016 October 28. Received 2016 October 25; in original form 2016 September 7.}

\begin{document}

\maketitle

\begin{abstract}

The origin of the low luminosity radio emission in radio-quiet AGN, is unknown. The detection of a positive correlation between the radio and X-ray emission would imply a jet-like origin, similar to that seen in `hard state' X-ray binary systems. In our previous work, we found no believable radio variability in the well known X-ray bright Seyfert 1 galaxy NGC~4051, despite large amplitude X-ray variability. In this study we have carefully re-analysed radio and X-ray observations using the same methods as our previous work, we again find no evidence for core radio variability. In direct contrast to our findings, another study claim significant radio variability and a distinctive anti-correlation between radio and X-ray data for the same source. The other study report only integral flux values and do not consider the effect of the changing array on the synthesised beam. In both our studies of NGC~4051 we have taken great care to account for the effect that the changing beam size has on the measured radio flux and as a result we are confident that our method gives more accurate values for the intrinsic core radio flux. However, the lack of radio variability we find is hard to reconcile because radio images of NGC~4051 do show jet-like structure. We suggest that the radio structures observed are likely the result of a previous period of higher radio activity and that the current level of radio emission from a compact nuclear jet is low.
\\

\end{abstract}


\section{Introduction}
\label{introking}


The origin of nuclear radio emission in the so-called `radio quiet'
AGN, particularly Seyfert galaxies, is still far from clear. The main
possibilities include emission from jets emanating from the central
black hole, emission from a central starburst or emission from some
sort of corona or wind \citep[e.g.][]{laor08}. Extended (on the order of a few
hundred parsec-scale) jet-like structure is seen in the nuclei of some
Seyfert galaxies \citep[e.g.][]{ulvestad05}. However, such structure
may be the relic of previous activity and does not necessarily mean
that a jet is currently active, i.e. that particles and energy are
currently being pumped in at the base of the jet. All the relevant radio
emission processes should produce a compact nuclear source. In some
cases \citep[e.g.][]{kharb14,bontempi12} VLBI observations reveal
sufficiently high brightness temperatures, coupled with flat or
inverted spectra, such that a jet is the only likely explanation.  However,
in many cases the brightness temperatures, even in VLBI
observations, are not high enough to rule out non-jet origins for the
radio emission and other diagnostics are necessary. 

The Seyfert 1 galaxy NGC~4051 is a case of particular interest. The
X-ray power spectral density (PSD) of this galaxy is very similar to
that of the X-ray binary (XRB) Cygnus X-1 when it is in a
`soft' X-ray spectral state \citep{McHardy05}. XRBs are found in a
variety of states with the two most common being the `hard' and `soft'
states. These states have been defined historically by their medium energy (2-10 keV)
X-ray spectra with the hard state having a higher proportion of high energy photons than the soft state.  
The X-ray PSD shapes of these two states are also quite different with the soft state only
showing one break from a low frequency slope of approximately $-1$ to a high frequency slope of $-2$. 
This break is also seen in hard state XRBs where a second break to a slope of $0$
is also seen at lower frequencies. For a detailed discussion of these X-ray states see \citet{mcclintock06}. 
Another major observational difference between the hard and soft states of XRBs is that an active radio
source is not usually found in the soft state XRBs. In these `soft' XRBs the jet appears to be `quenched' by the inner disc \citep{russell11,Tananbaum72}. The jets that are observed in the soft state (or in an intermediate state) XRBs are thought to be the result of `relic' extended emission from jets produced in a previous hard state \citep{Rushton12,Kalemci11}. Therefore, if an active radio source is found in NGC~4051, as opposed to a quenched jet, this would cast doubt on the currently accepted paradigm that Seyfert galaxies are the galaxy-scale analogues of `soft state' XRBs.

For the hard state XRBs (which have an active jet) there is a strong correlation between the X-ray and
radio fluxes \cite[e.g.][]{Hanni00,corbel03}, therefore, the detection of such a
correlation in Seyfert 1 AGN NGC~4051 would indicate an active jet in that source.

There have been two previous papers dealing with radio and X-ray
observations of NGC~4051. In \citet{Jones11}, hereafter referred to as Paper I, we presented 29
radio observations of NGC~4051 over the course of 16 months with
the Karl G. Jansky Very Large Array (hereafter referred to as the VLA).
This radio data was presented together with observations, every 2 days in X-rays with the Rossi X-ray Timing Explorer (RXTE).
The radio observations were made in all four VLA configurations
(A, B, C and D) including some of the combined array configurations (DnC, BnA and BnC). 
We analysed each of these radio observations very carefully in
order to remove contamination from unresolved extended emission whose
contribution is different in each configuration due to the different
size of the synthesised beam in each configuration.

In Paper I we found a faint flat spectrum radio core within the VLA A configuration with $\approx0.5$~mJy peak flux density at 8.4~GHz. We assume that A configuration provides the highest angular resolution and hence the least
contamination from unresolved extended emission. In fact, if we use the peak flux density value for the core at 1.65~GHz as measured in Chapter 4 of \citet{JonesThesis}, $F_{1.65~GHz}=0.24\pm0.03$~mJy, and assume the spectral index value for a flat spectrum source, $\alpha=-0.7$ in the relationship $S\propto\nu^{\alpha}$. Then we calculate a flux at 8.4~GHz for the core, using the VLBI, at $F_{8.4~GHz}\approx0.08$~mJy. Therefore, since our 8.4~GHz VLA A configuration peak flux is  $\approx0.4$~mJy brighter than the predicted VLBI value there may still be substantial unresolved extended emission. For the A configuration data points we found a very weak suggestion of positive X-ray/radio correlation and a very weak radio variation (of the order of $0.1$~mJy). However, we only have 6 A configuration data points in this work so this data alone is not sufficient data to draw any strong conclusions. From the full 16 month data set of 8.4~GHz data in all configurations 
(offset to A configuration) we could find no evidence of believable radio
variability, despite the very large amplitude X-ray variability for the same time period.
It should also be noted that the low amplitude intrinsic variability is more difficult to distinguish in the 
lower resolution/more compact B, C, D, DnC, BnA and BnC configurations. Also during these configurations the emission on the scale of the synthesised beam is dominated by the extended emission.

In a separate paper \citet{King11}, hereafter referred to as
K11, investigate the radio emission from NGC~4051 using 6 VLA 8.4~GHz radio observations from 4 different array configurations (A, B, BnC and C). K11 did not attempt to model the variation of the radio flux with changing array configuration and they do not take into account the contribution of the extended emission to the measured flux within the synthesised beam. Their X-ray data was from {\it Chandra}. K11 find an anti-correlation between the X-ray and radio fluxes for NGC~4051 (see top two panels in Figure~\ref{chandrawx} for the X-ray and radio measurements of K11).

Any interpretation of the origin of the
radio emission in NGC~4051 requires that we understand whether there is
any correlation, positive or negative or neither, between the X-ray and
radio emission. This paper is an effort to determine if there is an intrinsic correlation using a larger dataset. In Section 2 we present all the radio and X-ray datasets. In Section 3 we re-analyse all of the
8.4~GHz VLA observations of NGC~4051 from our earlier work (Paper I) and in that of \citet{King11}. We systematically take account of the contribution 
of extended emission in the various array configurations to produce an 8.4~GHz lightcurve which is offset to A configuration and investigate any possible variability.
In Section 4 we compare the resulting radio lightcurve, offset to A configuration, with the RXTE X-ray lightcurve. In this last section, using a larger dataset, we confirm the result of Paper I, i.e. that there is no believable long timescale (year long) radio variability found in NGC~4051 and no correlation between the X-ray and radio variability. The possible weak positive correlation on short (week long) timescales reported in Paper I is detected again in this work, but the radio observations are also consistent with being constant.  In Section 5 we state our conclusions and consider the implications of our results for the origin of the radio emission in NGC~4051.

\section{Radio Observations} 
\label{sec:obsRadio}

\subsection{8.4~GHz VLA Observations}
\label{sec:8.4}

Compact radio jets are associated with a flat radio spectrum whereas
the surrounding (often extended) radio emission has a steeper spectrum. Given that higher
frequency observations also provide better spatial resolution, high frequency observations (8.4~GHz) are preferred (over 4.8~GHz) for investigating jet emission in this study.
The 8.4~GHz data also provides the best compromise between angular resolution and sensitivity.

NGC~4051 was observed 37 times by the VLA at 8.4~GHz between 1991 and 2009 (see Table 1). 
These 37 observations are made up of 3 data sets referred to as the 1991, 2000/01 and 2008/09 datasets.
The 2000/01 data were previously discussed in Paper I and the 2008/9 data were first presented in K11.

The 1991 dataset is combined with the 2000/01 and 2008/09 datasets to investigate the 
long time scale variations of the nuclear flux density of NGC~4051 during A configuration. 
The 1991 dataset was taken between the 24th June and 1st September 1991.
The flux calibrator used was 3C286(1331+305) and the phase calibrator was FIRST J114915.3+393325. The phase calibrator was observed in
 between each observation of NGC~4051, for $\approx$~2~minutes on 3 occasions.

The 2000/01 radio data is presented in Table~1 and the reduction of this dataset was discussed previously in Section 2 of Paper I.

For the 2000/09 dataset NGC~4051 was observed a total of 6 times between 31st December 2008 (JD 245 4831.3) and 31st July 2009 (JD 245 5043.1) with a typical interval between observations of 10 weeks (compared with every 2 weeks for the 2000/01 data presented 
previously in Paper I). Observations were carried out in 2IF mode with 50~MHz bandwidth at both frequencies. 
During the observations the antennae switch from the source to the phase calibrator (12214+44114) every 3.33~minutes. 
The flux calibrator used was 3C286 (1331+305) which was observed for 3.33~minutes of integration time at the end of each
 observation (see Table ~\ref{obs} for more information).


\begin{center}

\begin{table}
   \caption[Observational details of VLA 1991, 2000/01 and 2008/09 data at 8.4~GHz.]
{Observational details of VLA 1991, 2000/01 and 2008/09 datasets at 8.4~GHz. The first set of data is from the Historical Very Large 
Array archive from 1991, and is referred to as the 1991 data set. The second set of data is as presented in Paper I,
and is referred to as the 2000/01 data set. 
The third dataset is Historical VLA/EVLA archival data which was presented in K11 and is re-reduced and anaylsed in this work; this is referred to as the 2008/09 data set.}
  \begin{tabular}{@{}lcc}
\\
     Date    & Configuration  & Time on Source\\
MJD & & (s) \\
&&\\
& \underline{1991} \\

     48431.000 & A  & 910\\
     48500.000 & A  & 3490 \\
&&\\
& \underline{2000/01}  \\

     51701.583 & C  & 1130\\ 

     51722.514 & DnC & 1170\\ 
     51734.542 & DnC & 1130\\ 
     51741.505 & DnC & 1090\\ 

     51750.467 & D & 1210\\ 
     51756.535 & D & 1210\\ 
     51771.465 & D & 1100\\ 
     51803.335 & D & 1140\\ 
     51817.342 & D & 770\\

     51852.222 & A & 1150\\ 
     51865.185 & A & 1160\\ 
     51879.148 & A & 1140\\ 
     51893.131 & A & 1080\\ 
     51909.066 & A & 1090\\ 
     51929.926 & A & 1140\\ 

     51941.001 & BnA & 1000\\ 
     51956.956 & BnA & 1070\\  

     51985.710 & B & 1130\\ 
     52005.781 & B & 1120\\ 
     52023.773 & B & 1120\\ 
     52034.660 & B & 1150\\ 
     52046.649 & B & 1120\\ 
     52060.660 & B & 1210\\ 

     52078.464 & BnC & 1250\\ 

     52091.505 & C & 1140\\ 
     52101.560 & C & 1120\\ 
     52114.444 & C & 770\\
     52132.351 & C & 1130\\ 
     52156.272 & C & 1230\\ 
&&\\
& \underline{2008/09} \\

     54831.814 & A  & 1820\\
     54899.672 & B & 1783\\
     54987.576 & BnC & 1873\\
     55005.693 & C & 1890\\
     55027.500 & C & 1230\\
     55043.589 & C & 1883\\
 \end{tabular}
\label{obs}
\end{table}
\end{center}

\subsection{Radio Analysis Method}
\label{sec:RAM}

\subsubsection{Image Reduction}
\label{sec:ImageReduction}

We reduced both the 1991 and 2008/09 VLA archive radio data using the Astronomical Image Processing Software (AIPS) following the same method that we used for the 2000/01 data in Paper I. 
In summary, our maps were made using the {\sc imagr} cleaning task within AIPS, we set clean fields around two fainter neighbouring sources, one approximately 5.5 arcmin to the NE and the other approximately 3.5 arcmins to the South. We standardised on 10,000 iterations of cleaning and no weighting was applied to any of the three datasets i.e. we did not restrict the {\sc uvrange} and we used a robust parameter$=0$ to produce all the images across the 1991, 2000/01 and 2008/09 datasets. No self calibration was applied to any of the images because the source was too faint. We used the task {\sc jmfit} to determine the flux density values of the core.
K11 use a slightly different reduction method, they use the Common Astronomy Software Application (CASA) package to reduce their images from 2008/09. They used the {\sc clean} algorithm to create their images using a threshold of 0.1~mJy, a Briggs weighting and a robust parameter$=0.5$. Like us, K11 do not perform any self calibration and they use a flux density calculation task which is equivalent to {\sc jmfit} which is called {\sc imfit} within CASA.

\subsubsection{Measuring Core Flux Density}
\label{sec:CoreBox}

The only part of NGC~4051 which can possibly vary on the few
months/year timescales is the very
innermost, unresolved, core and it is that which we are studying here. In order to restrict our radio flux measurement to
this core we restrict the box size within which we measure flux
so that it only encloses the core.

We therefore used the task {\sc cowindow} 
within {\sc aips} which allows the user to specify a box
size. We used the same box size for each image in the same array
configuration. The box size for each array configuration was chosen to
just encompass the synthesised beam for that configuration.We use the task {\sc jmfit} to measure accurately the peak and total/integral flux density of the core Gaussian components. Thus we used boxes of size $ 1''\times1''$, 
$2.55''\times1.69''$ and $8.4''\times8.4''$ for images from the A, B and C
configurations respectively, all centred on the position of peak flux. 

\begin{figure}
\includegraphics[width=0.45\textwidth,angle=270]{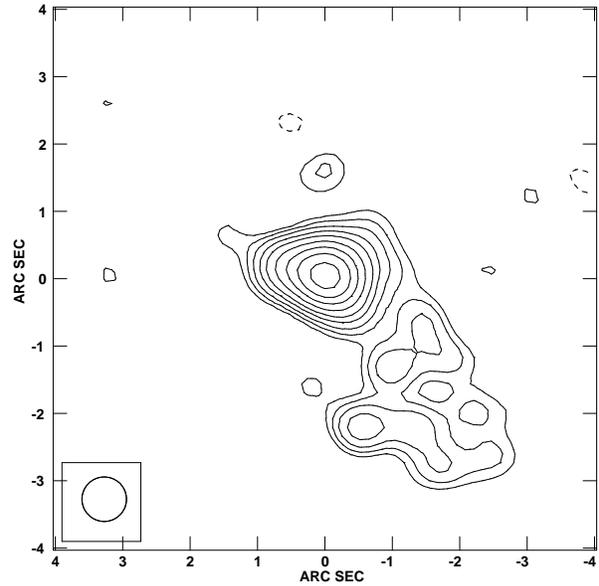}
\caption{ Image taken during B configuration from the 2008/09 dataset. The map is centred on the nucleus with North up and East to the left. Map made with a default restoring beam. The peak flux density from {\sc jmfit} is (10.1 $\pm$0.21 )$\times \rm{10^{-4}~Jy/beam}$. The rms noise level is  $2 \times \rm{10^{-05}~Jy}$ and contour levels are placed at rms noise $\times~-2.8, 2.8, 4,5.6, 8, 11, 16, 23, 32, 45, 64, 90, 127,180, 250 $. }
\label{Bmap}
\end{figure}

A map of NGC~4051 from the 2008/09 dataset during B array is shown in Figure~\ref{Bmap}, the extended structure is clearly visible to the South-East of the core. Note that the box size used to measure flux in this image, $2.55''\times1.69''$ when centred on the core includes all of the core flux while having only a small flux contribution from the extended structure.
The larger synthesised beams from the
more compact array configurations will, of course, include a larger
contribution from extended emission than in the more extended A
configuration. We take account of these differences using our
`offset' method, which we discuss in Section~\ref{sec:OffsetM}.

K11 used the CASA task {\sc imfit} to fit a Gaussian to the flux within a
box of size $8''\times 6''$ which they keep constant for
all 6 images from 4 different array configurations. They chose this
box size as they believe it will lead to
`inclusion of all structures present in the images', i.e. inclusion of
all radio emission from NGC~4051. However, our
extensive imaging (e.g. Paper I, \citet{Jones11} and \citet{JonesThesis}) shows
that there is extended emission on scales up to $20''$ and
that even if we select a constant box size, both the core and the
integral flux densities of any Gaussian fitted to the peak flux
density within that box will vary with array configuration as
different amounts of extended emission are included within the
`unresolved' core.

\subsubsection{Peak Flux Density vs Integral Flux Density}
\label{sec:peakvint}

In Paper I we showed that the peak flux density (offset to the
equivalent peak flux density that would have been obtained in
A-configuration observations) did not vary significantly during a 16
month period in 2000/2001. However the integral flux density of a
Gaussian fitted to the central source increased as
the synthesised beam increased. The peak flux density therefore
appears to be a better indicator of the radio luminosity of the
central engine. 

As another test of whether the peak or total flux density is the best
measurement of central core luminosity we compare, in Figure 4, our
measurements of the peak and integral flux densities to those presented in K11. Although we
use different box sizes, we agree on the peak flux densities. Thus the
peak provides a reasonably robust measurement.  

However, we do not agree on the integral flux densities. Our integral
flux densities increase in a predictable way as the synthesised beam
increases. For K11 the A-configuration integral flux density is
lowest, B is highest, and C is a little lower than B. In
A-configuration K11 resolve NGC~4051 into three components, as seen
previously, but do not resolve it in any other configuration.  Based
on their statement that they use a constant box size within which to
measure the flux density, we assume that the integral flux density which
they state is the sum of the fluxes of all Gaussian components within
that box. The integral flux density for A configuration is, however, noticeably lower than
that in any other configuration so K11 wonder whether some extended emission
may indeed be resolved out. However K11 find a lower integral
flux in their C configuration observations than in B-configuration and
so claim a real intrinsic decrease in core luminosity. We suggest an
alternative explanation related to the relative size of the
synthesised beam and the box within which one fits Gaussians to the
peak.

To investigate the relevance of chosen box size we
took the same $8''\times 6''$ box size as K11, then for the first
C configuration observation we measure an almost identical peak flux
density ($1.96\pm0.10$~mJy c.f. $2.22\pm0.08$~mJy for K11). Our total/integral flux
density is similar ($4.30\pm0.23$mJy) though not identical
($4.97\pm0.17$mJy for K11). If we halve the box size in each
dimension, the peak flux density does not change but the integral drops
by roughly $20\%$ to 3.4~mJy. If we double the box size in each dimension, again the peak flux density we measure is similar to K11 ($1.91\pm0.02$mJy) but the integral flux is about $11\%$ lower than the K11 integral flux density ($4.44\pm0.07$mJy). The reason for the drop in integral flux for the smaller box is that {\sc jmfit} fits a narrower Gaussian when the box is smaller. 

However, in B-configuration the K11 box size is significantly larger than the
beam and so the Gaussian fit is not restricted and flux is not
lost. This effect may make a small difference to the measured total
flux density.
We also note that when using the same box size as K11, we agree exactly on
the peak flux density for their B-configuration observations
($1.17\pm0.06$ mJy vs $1.13\pm0.04$ mJy by K11). However we measure 
$1.89\pm0.11$ mJy for the integral/total flux density compared to $5.99\pm0.20$mJy
by K11. In both cases, we let the Gaussians find their own width within the $8''\times 6''$ box.

We conclude that comparison of measurements of integral flux density
between different array configurations, for an unresolved source in
the presence of surrounding extended emission, is prone to a number of
uncertainties which make any measurement of variability
problematic. However the peak flux density is a more robust estimator
which will be agreed upon by all observers. As long as the different
contributions of extended emission in different array configurations
is understood, the peak flux density provides the best measurement of core luminosity.
Below, following Paper I, we repeat briefly the
'offset' method by which we take account of extended emission.

\begin{figure*}
\includegraphics[width=0.95\textwidth]{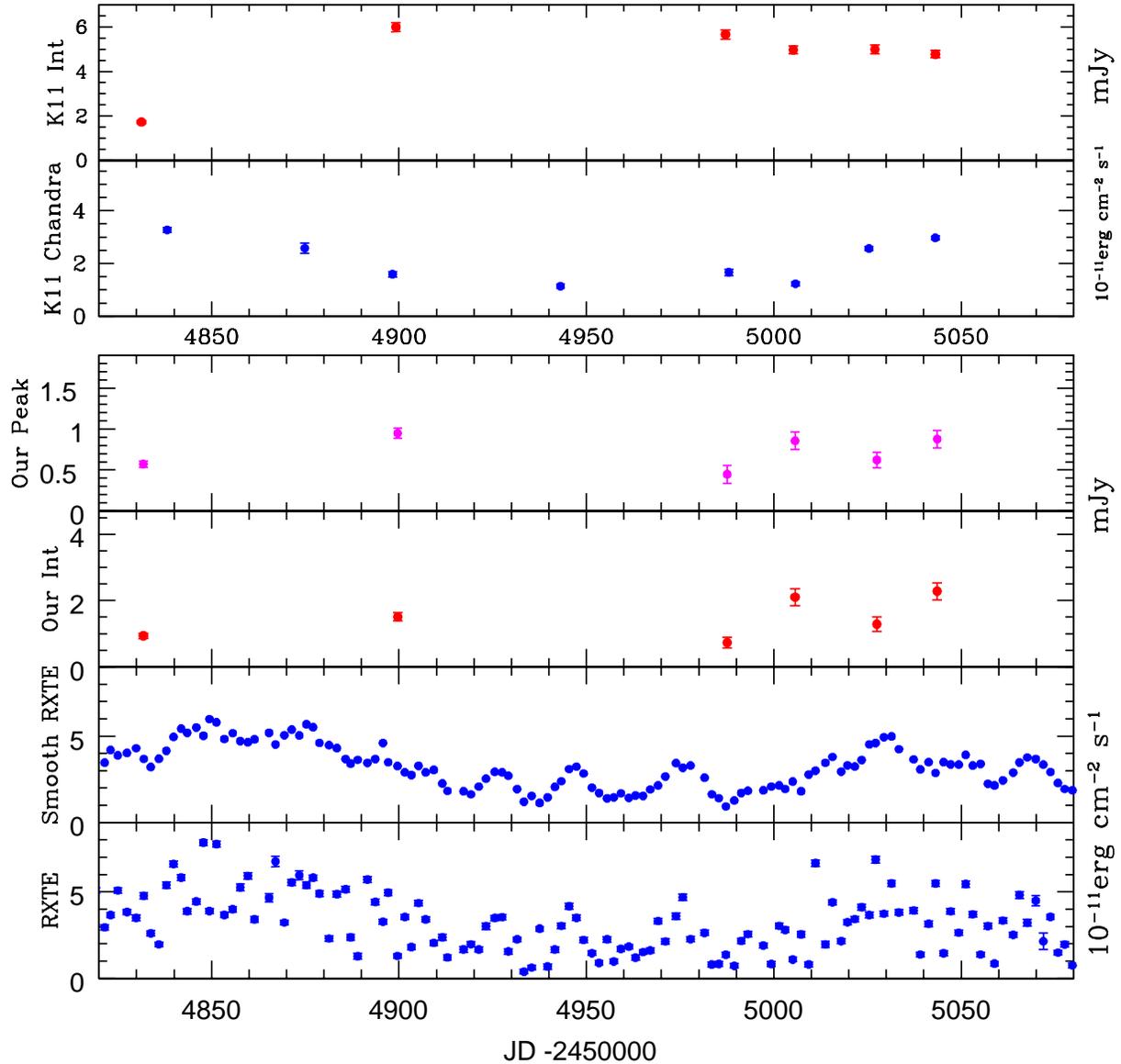}
\caption{Both the K11 radio and X-ray lightcurves and our radio and X-ray lightcurves plotted for period of the 2008/09 dataset. Top panel: The six integral flux density data points at $8.4~GHz$ presented in K11.  Second panel: The eight \emph{Chandra} X-ray points presented in K11. Third panel: Peak flux density at $8.4~GHz$, from our reduction of the 2008/09 data, offset to A configuration.  Fourth panel: Integral Intensity at $8.4~GHz$ from our reduction of the 2008/09 data offset to A configuration.  Fifth panel: The smoothed RXTE lightcurve for the 2008/09 time period. Sixth panel: The unsmoothed `observed' RXTE lightcurve.  All the X-ray data was integrated over $2-10~keV$ with units of ${10^{-11}\mathrm{~erg~cm^{-2}~s^{-1}}}$. All radio flux density measurements are in units of mJy. The long term variability of NGC~4051 detected by {\it Chandra} is consistent with the RXTE flux measurements from the same time period. The Spearman rank coefficient for the two X-ray lightcurves is 0.78, suggesting a high correlation between the two X-ray lightcurves on long timescales.}
\label{chandrawx}
\end{figure*}


\subsubsection{The `Offset' Method}
\label{sec:OffsetM}

Our extensive imaging over all array configurations (see particularly Chapter 4 of
\citet{JonesThesis}) shows structure over a wide range of
scale sizes. Changing array configuration changes the resolution and
the amount of extended emission within a synthesised beam and
thus alters both the peak and integral flux density associated with
the central source.

In the offset method, described originally in Paper I but summarised
here for convenience, we take for example, an
A-configuration observation and measure both the peak and integral
flux density associated with the central source. We then limit the 
the {\sc uvrange} in {\sc aips} so that the maximum UV limit, and resultant synthesised beam,
is the same as in B-configuration. We then again measure the peak and
integral flux densities associated with central source, which are both
larger than in the A-configuration. We then calculate the difference,
or offset, between the peak and integral flux densities as measured in
both configurations. In the above case we would measure a B to A offset, but
the same procedure is carried out for all arrays. Previously we have show that
the peak flux density provides a more accurate measurement of the true
intrinsic source luminosity than integral flux density so, from hereon, we only consider the peak flux density offset values. 
As there is more than one observation in any one configuration
we can measure the average offset and then from the spread, derive an
error in the offset. For the 2000/01 dataset (presented in Paper I) this resulted in a offset for A to B 
configuration equivalent to $0.32\pm0.07$~mJy. When offsetting over more than one configuration 
e.g. from C to B and then B to A configuration, we combined the offsets linearly and the errors in quadrature. 

To check the accuracy of the offset method we made a model of NGC~4051
by combining 28 clean components from a combined A, B, C and D dataset
(see Sections 4 and 5 of Paper I). We created 22 UV datasets from
the residual maps from the real data, and the 28 clean components. We
then performed the `offset method' on this `non varying' model data in
exactly the same way as we had done for the real dataset. The
resultant peak flux densities, offset to the A-configuration, should
be constant in all configurations. In Figure 11 of Paper I we show that
the resultant integral flux densities from all configurations, offset to
A configuration, are indeed constant to within an error
$\approx0.1$mJy. For the integral flux densities the offset method works
reasonably, but not as well as for the peak flux densities, showing
again that the peak flux densities provide a more reliable and
measurement of variation of core luminosity.

\begin{figure}
\begin{center}
\includegraphics[width=0.45\textwidth]{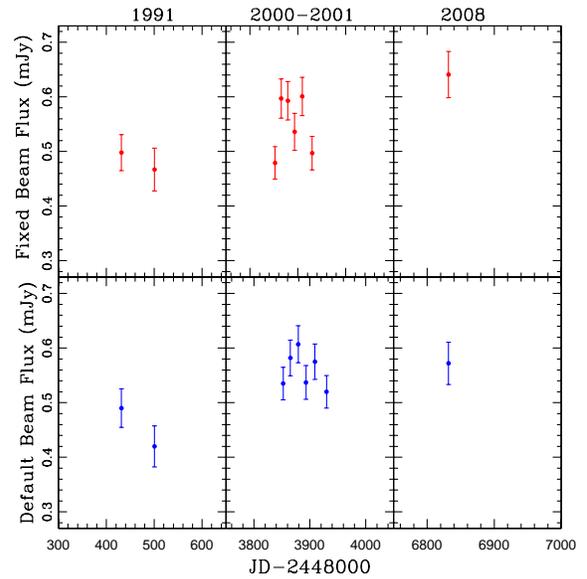}
\caption{A configuration radio peak flux values for the 1991, 2000/01 \& 2008 datasets (panels from left to right). The top panels with red points represent peak flux values from images made from the default beam size within the task {\sc imagr}. The bottom panels with blue points represent peak flux density values with a fixed beam size ($0.24''\times0.19''$ with PA~137~$^{\circ}$). The time axis is to scale within each panel, but date gaps exist between each of datasets/panels.}
\label{Araw}
\end{center}
\end{figure}
 
\subsection{Long Term Radio Variability within A-configuration}

In order to determine if there is any evidence for long term
variability in the core of NGC~4051 we first examined the A
configuration data as we consider this configuration to be the best
representation of the core flux as it is least affected by
contamination from extended emission. We analysed both the peak and
integral flux density values measured during the A configuration between 1991 and 2008 and 
combined the results with our previous analysis of the 2000/01 A configuration observations.

Figure~\ref{Araw} shows the variability of the nuclear core during A configuration only, at 8.4~GHz, for the 1991, 2000/01 and 2008. For all 3 data sets we have made images with both a fixed beam size (see top panels in graph) and the default beam size (see bottom panels in graph) respectively. For the fixed beam A configuration data the maps were remade with the mean restoring beam size $0.24''\times 0.19''$, a position angle PA of 137 degrees and robustness 0 (as in Paper I). For the default beam the restoring beam were left free, to be determined by the UV coverage and the use of robust 0 in {\sc imagr}. As in Paper I we used the task {\sc jmfit} to measure the peak and integral flux densities. This task fits a Gaussian model to the values within a box drawn around the source and outputs both the integral flux density (Jy) and the peak flux density (Jy/Beam). For all the flux density values quoted in this work the size of the fitted Gaussian component was left free. The quoted errors are the statistical error given by the {\sc jmfit} task, combined in quadrature with the $5\%$ error usually assumed as the maximum likely uncertainty on the flux density calibration.

We note that the peak flux density (particularly in the case of the default beam image) for the 2008 observation lies within the scatter of the values for the 2000/01 measurements. The corresponding values for the 1991 data lie a little below. We first fit a constant model to this data (with an intercept at the average flux density of 0.54~mJy). Taking the plotted errors at face value this model gives a reduced $\chi^{2} = 3.09$ for the fixed beam and a reduced $\chi^{2} = 2.50$ to the default beam.
If we allow for a model with a constant linear trend over all data we find that the gradient of the line is
$(2.12\pm1.00)\times 10^{-8}$ Jy/day with a reduced $\chi^{2}=2.15$ for the fixed beam data and is
$(2.18\pm0.78)\times 10^{-8}$ Jy/day with a reduced $\chi^{2}=1.35$ for the default beam data.
An F-test between the two models (constant and linear) has $F=4.49$
with a $P=0.07$ for the fixed beam data and $F=7.81$ with a $P=0.03$
for the default beam. This probability test indicates that the linear
model, ie a very slight increase in flux density with time, is a
better fit to the data than a constant model. However for both the
default and fixed beams the P-value is very close to 0.05, so the linear model is only marginally better, 
and the data can still be considered to be consistent with a constant i.e. no variability.

It is worth noting here that any variability, if present, is likely to
be more complex than a linear model so the above treatment is very
simplistic. Also, even after combining the {\sc jmfit} statistical 
error in quadrature with the standard $5\%$ error it is likely that we still underestimate the total error. For example, these errors do not
include any error associated with phase or amplitude calibration errors or side lobes from other sources which may not have been perfectly removed. The true total error is therefore hard to quantify but, in Paper I, we concluded that the actual error could be up to $30\%$ larger than the errors quoted previously. With these larger errors it is even more likely that the core flux density of NGC~4051 as measured within the A configuration is a constant, but, a slow increasing trend cannot be completely ruled out.

\section{Long Term Radio Variability Over All Array Configurations}
\label{sec:comp}

The main focus of the following sections is to offset the six 2008/09 radio flux values originally presented in K11 to an A configuration flux value so that they can be combined with our previous analysis of the 1991 and 2000/01 datasets. This combined lightcurve provides a comprehensive long term study of the core radio flux density of the AGN NGC~4051.

\subsection{Re-investigation of the 2008/09 radio data originally presented in K11}
\label{sec:kingradio}

\begin{figure*}
\begin{center}
\includegraphics[width=0.95\textwidth]{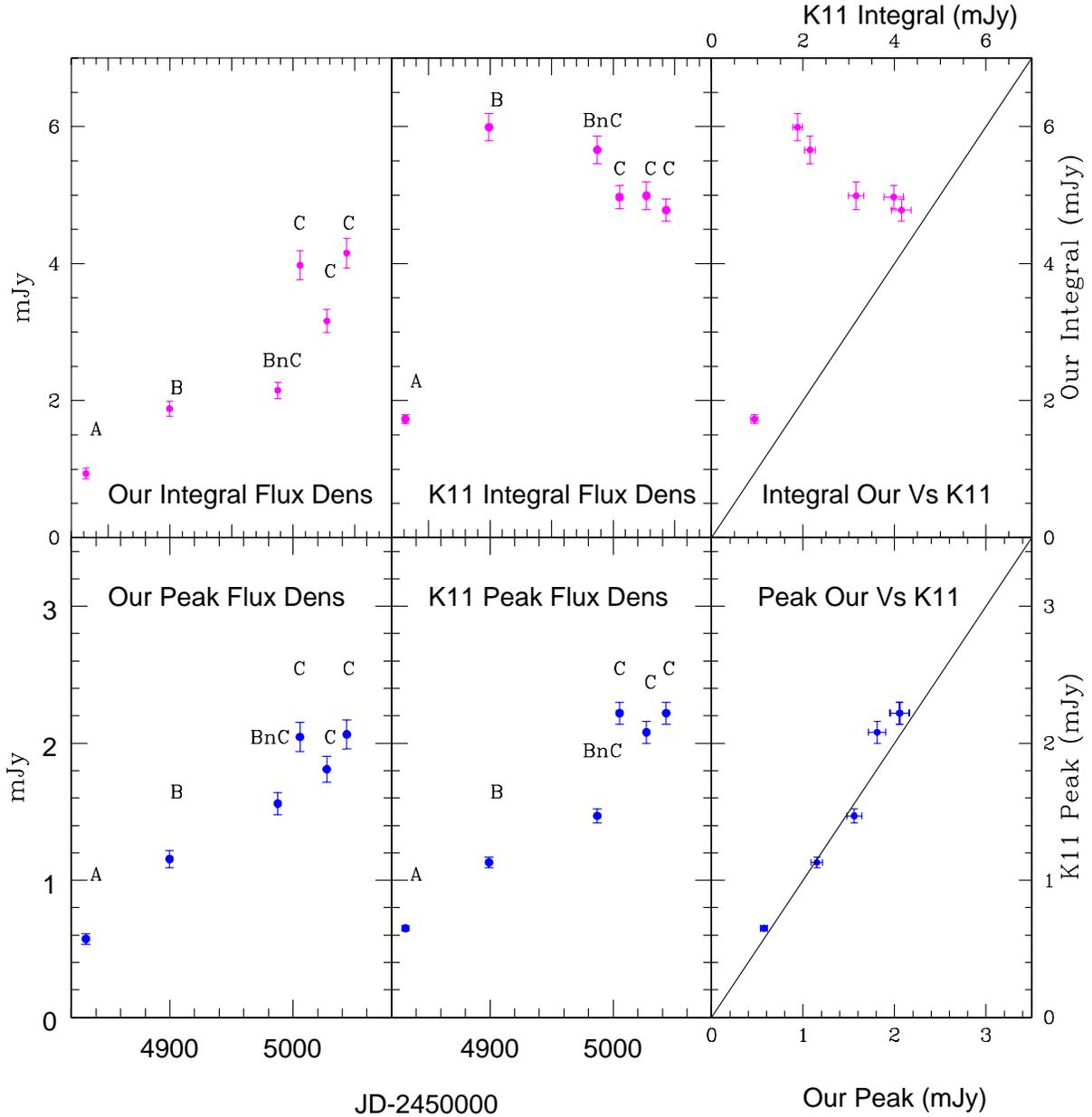}
\caption{Integral and peak flux density values are shown for both our analysis and the K11 analysis of 2008/09 data. While the lightcurves derived from the peak density measurements (bottom panels) are similar, the integral flux lightcurves (top panels) are dramatically different for the two methods. These differences can be observed from the plots on the far right which show our flux values versus the K11 flux values. Top left panel: Integral flux density values at 8.4~GHz from our reduction. Bottom left panel: Peak flux density values at 8.4~GHz from our reduction. Top middle panel: Integral flux densities at 8.4~GHz from K11. Bottom middle panel: Peak flux densities at 8.4~GHz from K11. Top right panel: Our Integral flux density values plotted against (Vs) the Integral flux density values from K11. Bottom right panel: Our Peak flux density values plotted against (Vs) the Peak flux density values from K11. All our maps were made with a default restoring beam.}
\label{rawmeking}
\end{center}
\end{figure*}

Here we compare our peak and integral radio flux density measurements (from our reduction of the 2008/09 data set) to that of K11. Figure~\ref{rawmeking} shows the peak and integral flux densities from our reduction on the left and the measurements of K11 in the middle panel. We then plot our peak flux versus the K11 peak measurements, and our integral flux versus K11 integral flux in the right hand panels. All values presented in this figure are the `raw' values, i.e. without any offset correction and are made from maps with a default restoring beam. K11 measure the total flux density within a $8''\times 6''$ box for all configurations. For reason stated previously in this work we used a different box sizes for each configuration, $ 1''\times1''$, $2.55''\times1.69''$ and $8.4''\times8.4''$ for A, B and C configurations respectively. 

The resultant peak and integral flux densities from both our work and that of K11 are shown, together with a comparison between the values derived from the two methods. The peak flux densities agree quite well (see bottom right box). However, the integral flux density values from the two studies are very different. It is very difficult to disentangle the reasons why our integrated fluxes are so different, and neither method for deriving integral flux densities is perfect, but based on points made previously in Section~\ref{sec:CoreBox} I suspect that K11 measured consistently higher integrated flux values due to the box size used.

\begin{center}
\begin{figure}
\includegraphics[width=0.45\textwidth]{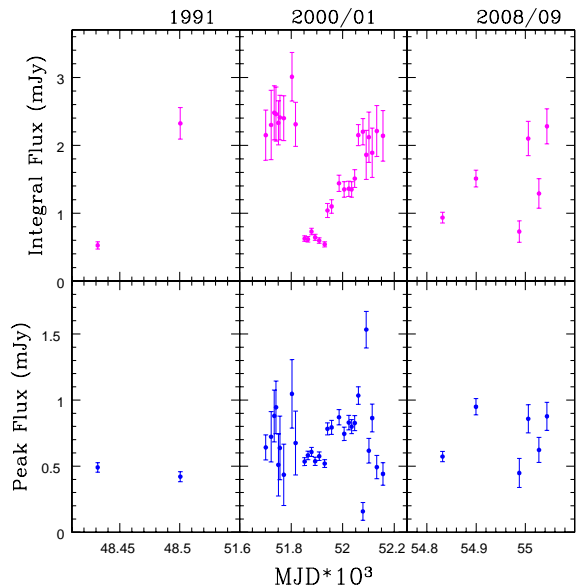}
\caption{Radio flux density values from 1991, 2000/01 and 2008/09 offset to A configuration. Core integral flux density (top panels) and peak flux density (bottom panels) at 8.4~GHz for 1991 on left, 2000/01 data in the middle and the 2008/09 data on the right. Flux densities for all observations are offset to the A configuration using the `offset method' discussed in Section~\ref{sec:OffsetM} and all images are made with a default restoring beam. For more details on which VLA configurations the data were in before the offset was applied see Table 1. The peak flux density is consistent with being constant i.e the majority of peak flux density values represent a factor of 0.5 variability (which represents $\approx0.5$~mJy change in flux.)}
\label{offsetking}
\end{figure}
\end{center}


\subsection{Long term investigation of ALL the 8.4~GHz radio data from 1991 to 2009.}
\label{sec:allradio}

In this section we produced an A configuration lightcurve that extends from 1991 to 2009. We have used our `offset method' on the 2008/09 dataset, combined this with the offset 2000/01 lightcurve that we produced earlier in Paper I and then finally used the `real' A configuration data from 1991. Note where the data is already in A configuration we did not apply the `offset method'. 

In Figure~\ref{offsetking} we show the 8.4~GHz peak and integral flux density lightcurve showing data from 1991 to 2009 offset to A configuration. The 1991 data (left panel) are `real' A configuration data. The middle panel shows the 2000/01 offset to A configuration data (from Paper I), and the right panels show the 2008/09 data offset to A configuration (from our reduction of the VLA archive data first presented by K11). All flux values were measured from maps made with a default restoring beam. 

From visual inspection of Figure~\ref{offsetking} the apparent variation seen in the 2008/09 data is within the scatter of the 2000/01 data, suggesting that the conclusions made in Paper I are still valid for this work, where we use a larger dataset over a longer time period. The peak flux density values presented in this figure are consistent with being constant given that the majority of peak flux density values represent a factor of 0.5 variability and it is likely that we have underestimated the errors. The majority of the peak variability data is represented by a $\approx 0.5$~mJy change in flux.

\section{The X-ray/Radio Relationship}

In the previous sections we have shown that there is no convincing
evidence for any variability in the radio flux density for the core of NGC~4051.
If the flux density is, in fact, constant then, given the large
amplitude X-ray variability, we would have to conclude that there is no
relationship between the X-ray and radio luminosities in NGC~4051.
However, the uncertainties in the radio flux densities are sufficiently
large so that we cannot rule out a small amount of radio variability.  In
this section, therefore, we look at the problem from a different angle.
We take the measured radio flux densities at face value and compare them
with X-ray fluxes taken at very close to the same time as the radio observations. We then search for
correlations between the two fluxes. 

Firstly we consider the `real' A configuration data (from 2000 to 2008) 
and compare with simultaneous X-ray observations. Next we comment on the findings of K11 for the 2008/09 dataset. Finally we examine all the radio data (from 2000 to 2009), which we offset to A configuration in Section~\ref{sec:allradio} and compare this with simultaneous X-ray data.

The X-ray data we use in this section consists of RXTE data and {\it Chandra} data. The RXTE data is from our long
term monitoring programme, and we also review the {\it Chandra} observations presented by K11 from 2008/09.

For the 2008/09 epoch X-ray lightcurves are available from both the RXTE and {\it Chandra} observatories (spanning from the 31st December 2008 (JD 245 4831.3) to 31st July 2009 (JD 245 5043.1)).
Both the smoothed and unsmoothed RXTE data and {\it Chandra} data as presented in K11 are shown in Figure~\ref{chandrawx}. The {\it Chandra} X-ray lightcurve is shown in the second panel down and the observed RXTE X-ray light curve is shown in the final, sixth panel. The RXTE data has a $\approx 2$~day sampling, compared with the {\it Chandra} data which is only sampled every $\approx 24$~days.

In both the {\it Chandra} and RXTE X-ray lightcurves the flux drops at about JD~245~4900 before beginning to rise again around JD~245~5000. However, for the duration of the lightcurves shown in Figure~\ref{chandrawx} the RXTE flux changes by a factor of 19.6 compared to a change by a factor of only 2.9 during the {\it Chandra} observations. Given the greater sampling of the RXTE data we give more weight to conclusions made using these data. It is impossible to detect the short time scale variations from the {\it Chandra} dataset because only six data points exist and these are separated by a range of $17-45$~days. Therefore, we suggest that only long timescale variations can be measured using the {\it Chandra} lightcurve.

Visual inspection of all of lightcurves presented in Figure~\ref{chandrawx} reveal no obvious correlation between the X-ray and radio. To produce a quasi-simultaneous dataset of X-ray and radio data (as in Paper I for 2000/01 data) we linearly interpolated the observed (unsmoothed) RXTE X-ray lightcurve to the exact dates of the 6 VLA 2008/09 observations at 8.4~GHz (see Table~\ref{xrayintking}). Due to the $\approx 2$~day sampling of the RXTE light curve interpolating between points results in a fairly accurate representation of the `real' X-ray flux at the time of interpolation, also the interpolated RXTE measurements are within the errors of the nearest observed flux values. In this work, we do not attempt to interpolate the {\it Chandra} data (reduced in K11) to the nearest radio dates because the separation in time between the 8.4~GHz radio and X-ray {\it Chandra} dates varies from 0.11-6.39~days (see Table~\ref{xrayintking}). If we were to linearly interpolate the {\it Chandra} data we would have to assume that the flux does not vary on short timescales and that it follows a simple straight line trend. However, this is not the case given the short timescale variability we observed in the RXTE lightcurve for the same epoch (sixth panel in Figure~\ref{chandrawx}). Note that the long term variability of the observed RXTE fluxes is consistent with the observed {\it Chandra} fluxes, we calculate a Spearman rank coefficient for the two X-ray lightcurves at 0.78, suggesting a high correlation between the two X-ray lightcurves on long timescales.

\begin{table*}
\linespread{1}
   \caption{8.4~GHz 2008/09 dataset observation dates presented with the nearest observed RXTE \& {\it Chandra} X-ray observation dates}

 $^{a}$~Dates are JD-2450000 \\
$^{b}$~ $\Delta$ time difference values are in units of days. \\
 $^{c}$~ X-ray fluxes are in units of $10^{-11}$ erg cm$^{-2}$ s$^{-1}$ and are shown as observed and
   interpolated to the time of the radio observation.\\
 $^{d}$~ {\it Chandra} observed values were taken 
directly from work by \citet{King11}.\\

  \begin{tabular}{@{}lcccccccc}
\\
\centering
     Radio    & Nearest    & $\Delta$(Radio & Nearest& $\Delta$(Radio- &Interpolated& Observed & Observed   \\ 
     Date  $^{a}$   &  RXTE Date$^{a}$ & -RXTE)$^{b}$ &{\it Chandra} Date$^{a}$ & -{\it Chandra})$^{b}$  & RXTE flux $^{c}$  &  RXTE flux  $^{c}$  &  {\it Chandra} flux  $^{ d}$\\
\hline 
     4831.81  & 4831.91    &  0.10  & 4838.20 & 6.39 & 4.71 $\pm$0.13  & 4.77$\pm$0.18    & 3.27 $\pm$0.09 \\ 
     4899.67  & 4899.59    &  0.08  & 4898.30 & 1.37 & 1.39 $\pm$ 0.12  & 1.30 $\pm$0.11   & 1.59 $\pm$0.10\\
     4987.58  & 4987.20    &  0.38  & 4988.00 & 0.42 & 1.26 $\pm$0.14   & 1.37 $\pm$0.14   & 1.67$\pm$0.12\\ 
     5005.69  & 5005.06    &  0.63  & 5005.80 & 0.11 & 1.51 $\pm$0.13   & 1.10 $\pm$0.13 & 1.23$\pm$0.08\\
     5027.50  & 5027.15    &  0.35  & 5025.40 & 2.10 & 6.38 $\pm$0.19  & 6.87 $\pm$0.19   & 2.57$\pm$0.07 \\ 
     5043.59  & 5043.17    &  0.42  & 5043.10 & 0.49 & 4.69 $\pm$0.16   & 5.49 $\pm$0.16   & 2.97$\pm$0.07\\ 
  \end{tabular}
\label{xrayintking}
\end{table*}

The $\approx2$~day sampling of the RXTE data is ideal for assessing the variability of the X-ray flux density in the nuclear region of NGC~4051. However, we have shown that measuring the radio variability of the core  accurately is much more challenging due to contamination from extended emission. It is important to mention here that the small number of radio data points (in comparison to the X-ray data points) and the poor time sampling of the radio lightcurves both greatly effect the conclusions made from the following analysis of a possible radio/X-ray correlation in NGC~4051. In the following sections we are discussing the peak flux density values from our own reduction of the 2000/01 and 2008/09 8.4~GHz unless otherwise stated.

\subsection{The relationship between RXTE X-ray data and the `real' A configuration radio data from 2000-2008}

As a first step in our investigation of a possible radio/X-ray correlation we used only A configuration radio data as we consider this to be the most free of systematic biases. Of the 6 data points at 8.4~GHz from 2008/09 only one A configuration point exists, from 2008. In order to investigate the long term correlation between the X-ray and radio emission observed during A configuration we therefore plot the peak flux density at 8.4~GHz data for both 2000/01 and 2008/09 data sets against the interpolated unsmoothed RXTE flux. Note that the 1991 radio data was taken before the launch of RXTE so we have no X-ray data for this epoch.

\begin{figure}
\begin{center}
\includegraphics[width=0.45\textwidth]{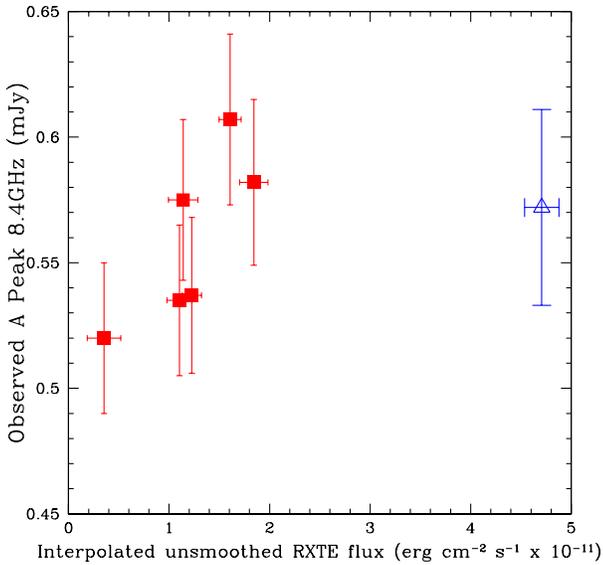}
\caption{Peak 8.4~GHz core flux densities from the `real' observed A configuration points (six from 2000/01 data and one from 2008). The 2000/01 data is shown as red solid squares and the 2008 data point is shown as a blue open triangle. A weak positive correlation is detected between the X-ray and radio data. We calculate a Spearman rank coefficient of 0.68 and a least squares fit gradient of $\approx~0.01$ with a reduced~$\chi^2=0.9$ calculated from 5 degrees of freedom (dof). This data is presented in Logarithmic space in the next figure.}
\label{realA}
\end{center}
\end{figure}

In Figure~\ref{realA} the radio/X-ray relationship for the period of the 2000 to 2008 data is shown. In this figure the red solid squares show the 2000/01 data and the blue open triangle is the single A configuration 2008 data point. Visual inspection of these seven points suggests a very weak positive radio/X-ray correlation between the observed flux values during A configuration. This weak positive correlation has a Spearman rank coefficient $\rho=0.68$ which for 7 data points is equivalent to a 1.7$\sigma$ detection. The least squares fit to this data has a gradient of $\approx0.01$ with a reduced~$\chi^2=0.9$ calculated from 5 degrees of freedom, which is consistent with being constant.

\begin{figure}
\begin{center}
\includegraphics[width=0.45\textwidth]{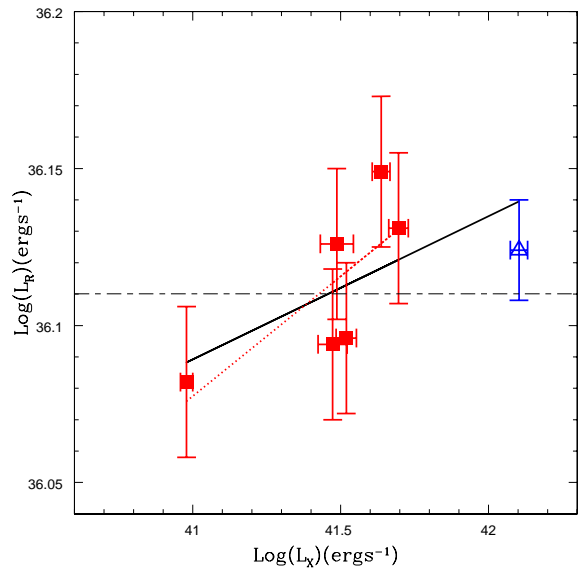}
\caption{Logarithmic relationship between the peak radio luminosity at 8.4~GHz versus interpolated unsmoothed X-ray luminosity of the `real' observed A configuration points. The six A configuration points from 2000/01 dataset are shown by solid red squares and fit with the red dotted line giving $\beta=0.08\pm0.03$ with a reduced $\chi^2=0.65$. The black solid line fits all seven points, including the one A configuration point from 2008 (shown as an open blue triangle). The least squares fit has a gradient of $\beta=0.05\pm0.03$ with a reduced $\chi^2=0.87$ calculated from five degrees of freedom (dof). The constant fit ($\beta=0$) is shown by a dotted and dashed line at the mean value and has a reduced $\chi^2=1.21$.}
\label{logA}
\end{center}
\end{figure}

To investigate the weak correlation observed in Figure~\ref{realA} we plot the relationship in logarithmic space in order to compare with the logarithmic relationship presented by other studies into the radio/X-ray correlation which exists for hard state objects e.g. \citet{corbel03},\citet{Gallo}, \citet{Gallo06} and in work by K11. In these studies of `hard state' objects (typically XRBs) the radio variability is generally described by $L_{R}\propto L_{X}^{\beta}$ where $\beta \approx0.7$. In Figure~\ref{logA} the logarithmic relationship between peak radio flux density (from the `real' A configuration points) and the interpolated unsmoothed X-ray luminosity at 2-10~keV is shown for NGC~4051. Data from the 2000/01 and 2008/09 datasets is fit with a least squares relationship to determine a value for the gradient, $\beta$ and a reduced~$\chi^2$ value. 

The six points from the 2000/01 epoch are shown as solid red squares and fit with a linear trend where $\beta=0.08\pm0.03$ (red dotted line) with and a reduced~$\chi^2=0.65$. The single A configuration point from 2008 is shown as a blue open triangle. The black solid line fits all 7 A configuration points and gives a least squares fit of $\beta=0.05\pm0.03$ again with a low reduced~$\chi^2=0.87$ calculated from 5 degrees of freedom (dof). So, even with this extra data point from 2008, we still find a very weak positive correlation between the radio and X-ray luminosities, with a maximum $\beta\approx0.08$, this is a factor of 8 shallower in logarithmic space than the $\beta\approx0.7$ correlation which exists for hard state objects. If we fit a constant ($\beta=0$) to all 7 data points at the mean value of $Log(L_{R})=36.11$ then the reduced $\chi^2=1.21$. Given that the reduced $\chi^2=0.87$ for the linear fit is closer to 1 we can say that the weakly positive trend is more likely than a constant model, but the difference is only marginal.

\subsection{The relationship between RXTE X-ray data and ALL our radio observations (offset to A configuration) from 2000 to 2009}

Next we investigated the possible weak positive correlation between the unsmoothed interpolated RXTE data and the peak 8.4~GHz `offset' lightcurve that spans from 2000 to 2009 (shown previously in Figure~\ref{offsetking}). If we again assume a simple linear trend relationship we measure a gradient=$0.018\pm0.008$ with a reduced $\chi^2=8.8$. All of the 2008/09 data values lie within the scatter of the more numerous 2000/01 data suggesting that a similar relationship between radio and X-ray emission exists at both epochs. The Spearman rank correlation co-efficient=$0.22$ which represents a 1.3$\sigma$ detection.The fact that the Spearman rank value (\textless~1) suggests that there is a lot of scatter in these data. This may be due to a weak correlation, but it could also be due to measurement errors that are large relative to the variations, or due to additional correlations with other parameters.

It is also worth noting here that one of the RXTE flux values observed in the 2008/09 epoch has an interpolated flux value which is a factor of $\approx 1.2$ higher than any of the RXTE values observed during the 2000/01 epoch. Similarly for the radio fluxes, one of the data points in the 2000/01 dataset has a value $\approx 1.5~$mJy, a factor of 2 higher than the mean radio flux (for the whole dataset) at $\approx 0.7$~mJy. These out lying points suggest `flare' events may have occurred at these times. 

\begin{figure}
\begin{center}
\includegraphics[width=0.45\textwidth]{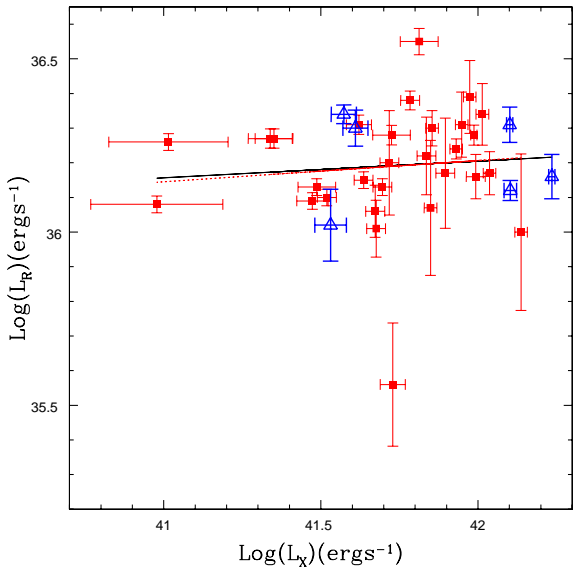}
\caption{Logarithmic relationship between peak 8.4~GHz core flux offset to A configuration and the unsmoothed interpolated 2-10~keV RXTE flux for the epoch of the 2000/01 and 2008/09 datasets. The 2000/01 data is shown as red squares where the dashed red line shows the linear fit to these points where $\beta= 0.12\pm0.02$ with reduced $\chi^2=5.8$ (from 28 dof's). The 2008/09 data set shown is shown as blue open triangles. The black solid line shows the fit to ALL the data and has a value of $\beta= 0.08\pm0.02$ with reduced $\chi^2=9.7$. The Spearman rank correlation co-efficient=$0.22$ and represents a 1.3$\sigma$ detection.}
\label{logALL}
\end{center}
\end{figure}

In Figure~\ref{logALL} we plot these data in logarithmic space to compare with past radio/X-ray correlation studies and calculate a $\beta$ value for this work. The figure shows the logarithmic relationship between the well sampled RXTE unsmoothed interpolated 2-10~keV data and the peak 8.4~GHz radio data, offset to A configuration.

The 2000/01 radio dataset, presented previously in Paper I, is shown as the red closed squares where the dashed red line shows the least squares fit to the 2000/01 data only with $\beta=0.12\pm0.02 $ with reduced $\chi^2=5.8$. The data from 2008/09 is shown as blue open triangles and the solid black line which is a fit to all the data has a value of $\beta=0.08\pm0.02$ with reduced $\chi^2=9.7$. The large reduced $\chi^2$ for both of these $\beta$ values shows that all fits are of poor quality. We suggest one reason for the poor quality of these fits is that the original errors, prior to offsetting to the A configuration, were likely an underestimation due to the fact that even in A configuration the core flux values are effected by extended features which are not resolved from the core. The extended emission itself is not varying, however, the variation to the core flux can vary between individual maps. If we assume these larger errors in radio flux then the quality of the fit (quantified by a reduced $\chi^2$) might improve. However, given that $\beta\approx0.08$ for the fit to the whole dataset we conclude that a weak positive correlation is more likely. If we assume that NGC~4051 has a jet then this measurement is much lower than the value for similar, hard state XRBs, where $\beta\approx0.7$. Our findings might suggest that parts of the radio emission from NGC~4051 are constant (not strongly varying) and are diluting the overall relationship.

\subsection{The distinct X-ray/radio relationship found by K11 for the 2008/09 epoch}
\label{distinct}

The lightcurves presented by K11 consist of eight {\it Chandra} X-ray observations and six VLA radio observations taken over a seven month period. Measurements in the X-ray and radio bands were carried out every two to four weeks (see Table 2 for observational details). Note that both the integral flux and peak flux density values are presented in work by K11, but only the integral flux value is used in their comparison with the X-ray lightcurve to measure any possible X-ray/radio correlation. The top two panels in Figure~\ref{chandrax} show the lightcurves we made from the X-ray and radio data presented in Table 2 and 3 of K11. Note that K11 do not use the second and fourth data points from the {\it Chandra} X-ray observations in their analysis because no `simultaneous' radio data existed for these X-ray points. Their final X-ray lightcurve shows variability by a factor of 3 and they calculate $\beta=-0.72 \pm 0.04$ for the ${\rm L_{R} \propto L_{X}^{\beta}}$ relationship. They also produce a second fit which excludes the only A configuration point, this fit is described by a weaker negative correlation with $\beta=-0.12 \pm 0.05$. The reason given by K11 for excluding the A configuration data point is that `this extended configuration has resolved out a substantial amount of flux in this high resolution image, which makes the flux value for this point inaccurately low'. However, in this work, we believe that the A configuration data is the most accurate representation of the core flux for a point source therefore it should not be excluded. 

In order to investigate if there is a time delay between the radio and X-ray emission K11 calculate the discrete cross correlation function (DCCF) for their radio and X-ray dataset (see Figure 9 in K11). After linearly interpolating the data every 17 days (which corresponded to the shortest time between two of the {\it Chandra} X-ray observations) they calculate a DCCF with a minimum of $-0.48 \pm 0.1$ at a time lag of $-2.5 \pm 5.3$~days, which they suggest indicates that X-ray dips are leading radio flares. Their predicted time lag value at $-2.5$~days at $\approx0.5\sigma$ is much less than their bin size of 17 days. Therefore, we suggest that their DCCF actually shows that X-ray and radio emission are constant in time and there is no lag detected.

\section{Summary and Conclusions}
\label{conc:king}

In Paper I we found, from 29 VLA observations taken over a 16 month
period in 2000/2001, that there was
no evidence for believable radio (8.4~GHz) variability in NGC~4051 and hence no evidence of a radio/X-ray
correlation. Subsequently K11, from 6 VLA radio observations in 2008/2009, claimed
a `distinctive' anti-correlation between the radio and X-ray
luminosities. For a relationship of the form $L_{R}\propto L_{X}^{\beta}$
we found $\beta=0.08\pm0.05$ for the 2000/01 dataset and K11 claimed 
$\beta=-0.72\pm0.04$ for the 2008/2009 observations.
For the `fundamental plane of black hole activity' $\beta\approx +0.7$ \citep{Merloni03plane,Gultekin09} so the $\beta$ value of K11
is completely inconsistent with the fundamental plane.

In this paper we have systematically reanalysed all of the VLA 8.4~GHz
radio observations, including those from Paper I, those presented in
K11 and additional archival observations from 1991. We have used our
`offset method' to offset each observation to equivalent A-configuration observations.
Following systematic analysis of this larger dataset we still find no
evidence for believable core radio variability from 1991 to 2009. For
the 2000/2001 and 2008/2009 observation epochs where we have high
quality RXTE X-ray observations, we find $\beta=0.05\pm0.03$, and hence no evidence for 
the `distinctive' anti-correlation, relationship proposed in K11 between the
radio and X-ray fluxes.

One obvious explanation of the lack of detectable radio variability is that the
large majority of the core radio emission does not come from the
compact jet but comes either fro[e.g.][]m relic radio emission from a
previously active core, from a wind or corona or from a starburst
region. However jet-like structures are clearly seen, on scales of
tens of pcs, in VLBI imaging \citep{GiroVLBI} with a size less than a few
pc. Although larger scale emission, which would not be resolved in VLA
A configuration observations, would not be able to vary on timescales
less than tens of years ($1''$ corresponds to a light crossing
timescale of about 240 years at the 15.2 Mpc distance of NGC~4051
\citep{Russell}), the light crossing time across the unresolved VLBI
core \citep{GiroVLBI} is only a few months. Considering the
evidence for jet emission in radio imaging of NGC~4051, the lack of
large amplitude radio variability is at first sight somewhat puzzling
given that jet-dominated quasars vary by far greater amounts
(e.g. factor of 5 variability seen in 4C~29.45 \citep{Hovatta08}).

One possibility is that the lack of variability may be an orientation
effect if we are viewing the jet side on.  Another possibility is that
there is real radio variability, but the variable component is
swamped by unresolved constant emission from larger scale regions and
so is not detectable. For example, we find from Figure~\ref{chandrawx} that the X-rays vary by a factor of $10$, while the
variability in the radio flux density is about 0.1~mJy at 8.4~GHz (equivalent to a factor of 0.3 variability). In
order to have a radio/X-ray relation following a power law with an
index of 0.7, we need to have an 8.4~GHz radio flux density that ranges
from about 0.02 to 0.12. The core VLBI flux density from \citet{GiroVLBI}
 is 0.24~mJy at 1.65~GHz. If the spectrum is optically
thin in the few GHz range, or there is still some extended,
non-variable emission contributing at 8.4~GHz, then this flux density
may be consistent with the upper end of the radio flux density range
needed to fit this source onto a standard (i.e. $\beta=0.7$) radio/X-ray
correlation.

Another possibility is that NGC~4051 does not have a jet, and its lack of radio variability is explained by radio emission originating in the corona rather than in the jet, see Section~6.2 of \citet{Jones11}. However, this possibility cannot explain the collimated jet structure which is detected in \citet{GiroVLBI}, chapter 4 of \citet{JonesThesis} and \citet{mch04}. 

Another factor which may hinder the discovery of the real radio/X-ray
relationship in NGC~4051 is the rapid X-ray variability and lack of
simultaneous X-ray and radio observations.
In a similar study by \citet{Bell11} on the LINER NGC~7213, which has
a black hole 40 times heavier than that in NGC~4051, the X-ray
emission leads the 8.4~GHz radio emission by $24\pm12$~days. If the
relationship between the X-ray and radio emission in NGC~4051 follows a
similar pattern to that in NGC~7213 (which, of course, is unknown at
present), then scaling by mass we would expect an X-ray/radio lag of roughly
$0.5$~days in NGC~4051. However the X-ray emission from NGC~4051
varies by factors $>2$ on timescales of a few thousand seconds e.g.
\citet{McHardy,Elme2010,Jones11,Vaughan11}. Thus lack of
simultaneity between radio and X-ray observations would introduce
substantial noise source into their correlation and the time separation between the measurements of the X-ray and radio from the core should not be separated by timescales of 2-6 days, which is the case for the K11 {\it Chandra}/8.4~GHz data.

We conclude that there is no evidence within current VLA observations for
radio core variability in NGC~4051, but radio variability may well be
occurring on sub-pc scales which is diluted by extended constant
emission which is not resolved by the VLA. It would, however, be
possible to detect such variability with VLBI observations. Hence a
relatively intensive VLBI and X-ray monitoring campaign could
determine whether NGC~4051, and by extension other Seyfert galaxies, do
contain synchrotron emitting jets similar to those found in hard state
X-ray binaries or whether their radio emission is not jet-related.

\vspace{5mm}
{\bf ACKNOWLEDGMENTS}\\

We thank David Williams, Juan Hernandez, Sam Connolly, James Matthews, Elme Breedt, Evan Keane, Phil Uttley, Derek Moss, Eric Greisen, Rob Fender, Grace Thomson, Anthony Rushton, Julie Jones and an anonymous referee for useful discussions and comments. IMcH acknowledges support under STFC ST/M001326/1. All data used in this work is freely available from the VLA, RXTE and {\it Chandra} archives.

\bibliographystyle{mn2e} 
\bibliography{imh,refs} 

\end{document}